\documentclass[twocolumn,letterpaper,aps,prl,superscriptaddress,showpacs,floatfix,preprintnumbers,longbibliography]{revtex4-1}

\usepackage{graphicx}   
\usepackage{xspace}     
\usepackage{url}
\usepackage[mathlines]{lineno}
\usepackage{bm}
\usepackage{amsmath}
\usepackage{xcolor}
\usepackage[driverfallback=dvipdfm]{hyperref}
\usepackage{ulem}
\hypersetup{
	colorlinks = true,
	citecolor= blue,
	linkbordercolor = cyan
}
\newcommand{\pt}{\mbox{$p_T$}\xspace}

\newcommand{\sqsn}{\mbox{$\sqrt{s_{_{NN}}}$}\xspace}

\newcommand{\lam}{\mbox{$\Lambda$}\xspace}
\newcommand{\alam}{\mbox{$\bar{\Lambda}$}\xspace}

\newcommand{\smath}[1]{\texorpdfstring{#1}{bbb}}

\def \be {\begin{equation}}
\def \ee {\end{equation}} 
\def \bea {\begin{eqnarray}}
\def \eea {\end{eqnarray}}

\def \snn {$\sqrt{s_{_{NN}}}$\xspace} 
\usepackage{color}
\definecolor{orange}{cmyk}{0.,0.353,1.,0.}    

\usepackage{lineno}
\newcommand*\patchAmsMathEnvironmentForLineno[1]{
  \expandafter\let\csname old#1\expandafter\endcsname\csname #1\endcsname
  \expandafter\let\csname oldend#1\expandafter\endcsname\csname end#1\endcsname
  \renewenvironment{#1}
     {\linenomath\csname old#1\endcsname}
     {\csname oldend#1\endcsname\endlinenomath}}
\newcommand*\patchBothAmsMathEnvironmentsForLineno[1]{
  \patchAmsMathEnvironmentForLineno{#1}
  \patchAmsMathEnvironmentForLineno{#1*}}
\AtBeginDocument{
\patchBothAmsMathEnvironmentsForLineno{equation}
\patchBothAmsMathEnvironmentsForLineno{align}
\patchBothAmsMathEnvironmentsForLineno{flalign}
\patchBothAmsMathEnvironmentsForLineno{alignat}
\patchBothAmsMathEnvironmentsForLineno{gather}
\patchBothAmsMathEnvironmentsForLineno{multline}
}

\begin{document}
%

\title{Hyperon polarization along the beam direction relative to the second and third harmonic event planes in isobar collisions at \smath{\snn = 200 GeV}}

\affiliation{Abilene Christian University, Abilene, Texas   79699}
\affiliation{Alikhanov Institute for Theoretical and Experimental Physics NRC "Kurchatov Institute", Moscow 117218}
\affiliation{Argonne National Laboratory, Argonne, Illinois 60439}
\affiliation{American University in Cairo, New Cairo 11835, Egypt}
\affiliation{Ball State University, Muncie, Indiana, 47306}
\affiliation{Brookhaven National Laboratory, Upton, New York 11973}
\affiliation{University of Calabria \& INFN-Cosenza, Rende 87036, Italy}
\affiliation{University of California, Berkeley, California 94720}
\affiliation{University of California, Davis, California 95616}
\affiliation{University of California, Los Angeles, California 90095}
\affiliation{University of California, Riverside, California 92521}
\affiliation{Central China Normal University, Wuhan, Hubei 430079 }
\affiliation{University of Illinois at Chicago, Chicago, Illinois 60607}
\affiliation{Creighton University, Omaha, Nebraska 68178}
\affiliation{Czech Technical University in Prague, FNSPE, Prague 115 19, Czech Republic}
\affiliation{National Institute of Technology Durgapur, Durgapur - 713209, India}
\affiliation{ELTE E\"otv\"os Lor\'and University, Budapest, Hungary H-1117}
\affiliation{Frankfurt Institute for Advanced Studies FIAS, Frankfurt 60438, Germany}
\affiliation{Fudan University, Shanghai, 200433 }
\affiliation{University of Heidelberg, Heidelberg 69120, Germany }
\affiliation{University of Houston, Houston, Texas 77204}
\affiliation{Huzhou University, Huzhou, Zhejiang  313000}
\affiliation{Indian Institute of Science Education and Research (IISER), Berhampur 760010 , India}
\affiliation{Indian Institute of Science Education and Research (IISER) Tirupati, Tirupati 517507, India}
\affiliation{Indian Institute Technology, Patna, Bihar 801106, India}
\affiliation{Indiana University, Bloomington, Indiana 47408}
\affiliation{Institute of Modern Physics, Chinese Academy of Sciences, Lanzhou, Gansu 730000 }
\affiliation{University of Jammu, Jammu 180001, India}
\affiliation{Joint Institute for Nuclear Research, Dubna 141 980}
\affiliation{Kent State University, Kent, Ohio 44242}
\affiliation{University of Kentucky, Lexington, Kentucky 40506-0055}
\affiliation{Lawrence Berkeley National Laboratory, Berkeley, California 94720}
\affiliation{Lehigh University, Bethlehem, Pennsylvania 18015}
\affiliation{Max-Planck-Institut f\"ur Physik, Munich 80805, Germany}
\affiliation{Michigan State University, East Lansing, Michigan 48824}
\affiliation{National Research Nuclear University MEPhI, Moscow 115409}
\affiliation{National Institute of Science Education and Research, HBNI, Jatni 752050, India}
\affiliation{National Cheng Kung University, Tainan 70101 }
\affiliation{The Ohio State University, Columbus, Ohio 43210}
\affiliation{Panjab University, Chandigarh 160014, India}
\affiliation{NRC "Kurchatov Institute", Institute of High Energy Physics, Protvino 142281}
\affiliation{Purdue University, West Lafayette, Indiana 47907}
\affiliation{Rice University, Houston, Texas 77251}
\affiliation{Rutgers University, Piscataway, New Jersey 08854}
\affiliation{University of Science and Technology of China, Hefei, Anhui 230026}
\affiliation{South China Normal University, Guangzhou, Guangdong 510631}
\affiliation{Sejong University, Seoul, 05006, South Korea}
\affiliation{Shandong University, Qingdao, Shandong 266237}
\affiliation{Shanghai Institute of Applied Physics, Chinese Academy of Sciences, Shanghai 201800}
\affiliation{Southern Connecticut State University, New Haven, Connecticut 06515}
\affiliation{State University of New York, Stony Brook, New York 11794}
\affiliation{Instituto de Alta Investigaci\'on, Universidad de Tarapac\'a, Arica 1000000, Chile}
\affiliation{Temple University, Philadelphia, Pennsylvania 19122}
\affiliation{Texas A\&M University, College Station, Texas 77843}
\affiliation{University of Texas, Austin, Texas 78712}
\affiliation{Tsinghua University, Beijing 100084}
\affiliation{University of Tsukuba, Tsukuba, Ibaraki 305-8571, Japan}
\affiliation{University of Chinese Academy of Sciences, Beijing, 101408}
\affiliation{Valparaiso University, Valparaiso, Indiana 46383}
\affiliation{Variable Energy Cyclotron Centre, Kolkata 700064, India}
\affiliation{Wayne State University, Detroit, Michigan 48201}
\affiliation{Yale University, New Haven, Connecticut 06520}

\author{M.~I.~Abdulhamid}\affiliation{American University in Cairo, New Cairo 11835, Egypt}
\author{B.~E.~Aboona}\affiliation{Texas A\&M University, College Station, Texas 77843}
\author{J.~Adam}\affiliation{Czech Technical University in Prague, FNSPE, Prague 115 19, Czech Republic}
\author{J.~R.~Adams}\affiliation{The Ohio State University, Columbus, Ohio 43210}
\author{G.~Agakishiev}\affiliation{Joint Institute for Nuclear Research, Dubna 141 980}
\author{I.~Aggarwal}\affiliation{Panjab University, Chandigarh 160014, India}
\author{M.~M.~Aggarwal}\affiliation{Panjab University, Chandigarh 160014, India}
\author{Z.~Ahammed}\affiliation{Variable Energy Cyclotron Centre, Kolkata 700064, India}
\author{A.~Aitbaev}\affiliation{Joint Institute for Nuclear Research, Dubna 141 980}
\author{I.~Alekseev}\affiliation{Alikhanov Institute for Theoretical and Experimental Physics NRC "Kurchatov Institute", Moscow 117218}\affiliation{National Research Nuclear University MEPhI, Moscow 115409}
% ...

\collaboration{STAR Collaboration}\noaffiliation

\date{\today}

\begin{abstract} 
The polarization of \lam and \alam hyperons along the beam direction has been measured relative to the second and third harmonic event planes in isobar Ru+Ru and Zr+Zr collisions at \snn = 200 GeV.
This is the first experimental evidence of the hyperon polarization by the triangular flow originating from the initial density fluctuations.
The amplitudes of the sine modulation for the second and third harmonic results are comparable in magnitude, increase from central to peripheral collisions, and show a mild \pt dependence. 
The azimuthal angle dependence of the polarization follows the vorticity pattern expected due to elliptic and triangular anisotropic flow, and qualitatively disagree with most hydrodynamic model calculations based on thermal vorticity and shear induced contributions.
The model results based on one of existing implementations of the shear contribution lead to a correct azimuthal angle dependence, but predict centrality and \pt dependence that still disagree with experimental measurements. Thus, our results provide stringent constraints on the thermal vorticity and shear-induced contributions to hyperon polarization.
Comparison to previous measurements at RHIC and the LHC for the second-order harmonic results shows little dependence on the collision system size and collision energy.
\end{abstract}

\maketitle


\setlength\linenumbersep{0.10cm}

%
The observation of the \lam hyperon polarization in heavy-ion collisions~\cite{STAR:2017ckg,STAR:2018gyt,Adam:2019srw,ALICE:2021pzu} opens new directions in the study of the fluid and spin dynamics.
The global polarization is understood to be a consequence of the partial conversion of the orbital angular momentum of colliding nuclei into the spin angular momentum of produced particles via spin-orbit coupling~\cite{Liang:2004ph,Voloshin:2004ha,Becattini:2007sr} analogous to the Barnett effect~\cite{Barnett:1916,Barnett:1935}. 
Its observation characterizes the system created in heavy-ion collision as the most vortical fluid known~\cite{STAR:2017ckg}.
Recent measurements with $\Xi$ and $\Omega$ hyperons~\cite{STAR:2020xbm} confirm the fluid vorticity and global polarization picture of heavy-ion collisions.

\begin{figure}[hbt]
\begin{center}
\includegraphics[width=\linewidth]{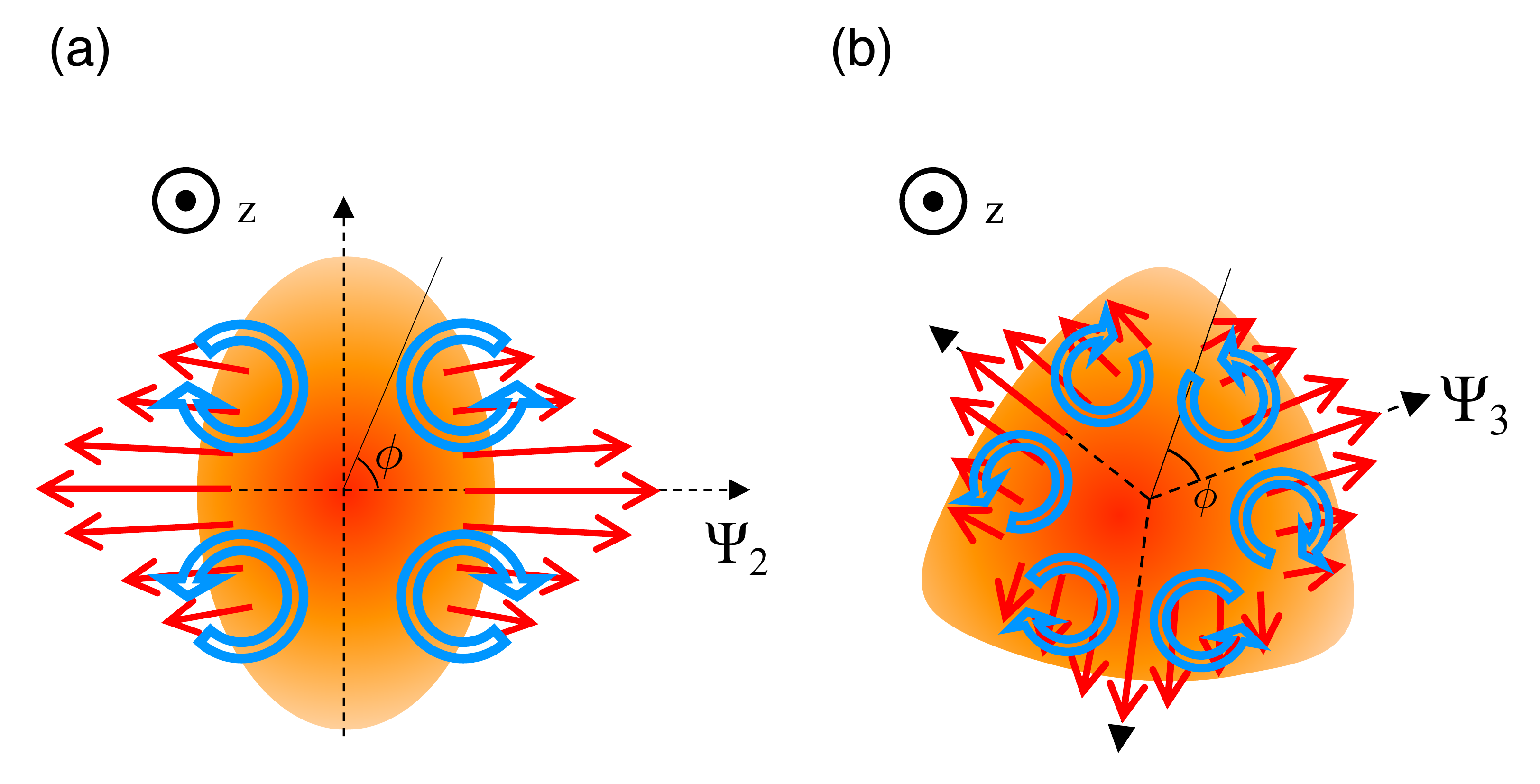}
\caption{Sketches illustrating the initial geometry, (a) elliptical shape and (b) triangular shape, viewed from the beam direction in heavy-ion collisions. Solid arrows denote flow velocity indicating stronger collective expansion in the direction of the event plane angle $\Psi_n$; open arrows indicate vorticities.}
\label{fig:cartoon}
\end{center}
\end{figure}

In non-central heavy-ion collisions, the initial geometry of the system in the transverse plane has roughly an elliptical shape as depicted in Fig.~\ref{fig:cartoon}(a). 
The difference in pressure gradients in the directions of the shorter and longer axes of the ellipse leads to preferential particle emission into the shorter axis, a phenomenon known as elliptic flow. 
Expansion velocity dependence on the azimuthal angle leads to generation of the vorticity component along the beam direction and therefore particle polarization~\cite{Voloshin:2017kqp,Becattini:2017gcx}.
\lam hyperon polarization along the beam direction due to elliptic flow was first observed in Au+Au collisions at \snn = 200 GeV by the STAR experiment~\cite{Adam:2019srw} and later in Pb+Pb collisions at \snn = 5.02 TeV by the ALICE experiment~\cite{ALICE:2021pzu}.
Sometimes such polarization driven by anisotropic flow is referred to as ``local polarization"~\cite{Gao:2012ix,Xia:2018tes}. 

While various hydrodynamic and transport models~\cite{Karpenko:2016jyx,Li:2017slc,Sun:2017xhx,Xie:2017upb,Ivanov:2019ern,Vitiuk:2019rfv} are able to describe the energy dependence of the global polarization reasonably well, most of them predict an opposite sign for the beam direction component of the polarization and greatly overpredict its magnitude~\cite{Becattini:2017gcx,Xia:2018tes,Fu:2020oxj,Florkowski:2019voj}.
On the other hand, the calculations based on a simple blast-wave model~\cite{Schnedermann:1993ws,Retiere:2003kf} utilizing only kinematic vorticity and without the temperature gradient and acceleration contributions can describe the data well~\cite{Adam:2019srw}. 
This situation has been referred to as the ``spin puzzle" challenging the understanding of the fluid and spin dynamics in heavy-ion collisions.
Recently, the inclusion of the shear-induced polarization (SIP) in addition to the thermal vorticity was proposed to help describing the experimental results on the  polarization along the beam direction~\cite{Fu:2021pok,Becattini:2021iol}.
However, these calculations strongly depend on the implementation details of the shear contributions~\cite{Alzhrani:2022dpi}. 
Furthermore, the shear-induced contribution may not be enough to fully understand the data~\cite{Florkowski:2021xvy} and the ``spin puzzle" remains to be resolved.

As predicted in Ref.~\cite{Voloshin:2017kqp}, in addition to the elliptic-flow-induced polarization, the higher harmonic flow ~\cite{PHENIX:2011yyh,ALICE:2011ab,ATLAS:2012at,CMS:2012xss,STAR:2013qio} originating from the initial density fluctuations should also induce vorticity and polarization. 
Figure~\ref{fig:cartoon}(b) depicts a triangular-shape initial condition with vorticity components along the beam direction induced by triangular flow characterized by its reference angle ($\Psi_3$).
The resulting polarization would have different shear-induced contribution than that for the elliptic-flow-induced polarization, thus providing unique constraints on the shear-induced contributions. Furthermore such a vorticity might depend on the system size~\cite{Voloshin:2017kqp}.
It is of great interest to investigate whether such a complex vorticity is indeed created.
More experimental data, especially from different collision systems and with respect to higher order event planes, are awaited for a better understanding of the local polarization phenomenon and to better constrain theoretical models.

In this Letter, we present \lam  and \alam hyperon polarization along the beam direction relative to the second-order event plane, and, for the first time, to the third-order event plane in isobar Ru+Ru and Zr+Zr collisions at \snn = 200 GeV. 
The high statistics and excellent quality isobar data taken by STAR for the chiral magnetic effect search~\cite{STAR:2021mii} provide a great opportunity for polarization studies in collisions of smaller nuclei compared to Au+Au, as well as to study polarization due to higher harmonic anisotropic flow.
The measurements are performed as a function of collision centrality and hyperon transverse momentum. 
The results are compared to hydrodynamic model calculations as well as to the previous second-order event plane measurements at RHIC and the LHC.

The data of isobar Ru+Ru and Zr+Zr collisions at \sqsn = 200~GeV were collected in 2018 with the STAR detector. 
Charged-particle tracks were reconstructed with the time projection chamber (TPC)~\cite{tpc} covering the full azimuth and a pseudorapidity range of $|\eta|<1$.
The collision vertices were reconstructed using the measured charged-particle tracks and were required to be within $(-35, 25)$~cm relative to the TPC center in the beam direction. The asymmetric cut was applied to maximize the statistics since the vertex distribution became asymmetric due to online vertex selection~\cite{STAR:2021mii}.
The vertex in the radial direction relative to the beam center was required to be within 2~cm to reject background from collisions with the beam pipe.  
Additionally, the difference between the vertex positions along the beam direction from the vertex position detectors (VPD)~\cite{vpd} located at forward and backward pseudorapidities ($4.24<|\eta|<5.1$) and that from the TPC was required to be less than 5~cm to suppress pileup events. 
In order to further suppress the out-of-time pileup events, the events with large difference between the total number of the TPC tracks and the number of the tracks matched with a hit in the time-of-flight (TOF) detector~\cite{tof} were also removed.
Quality assurance based on the event quantities that reflect the detector performance changing with time was performed following the study in Ref.~\cite{STAR:2021mii}.
These selection criteria yielded about 1.8 (2.0) billion minimum bias good events for Ru+Ru (Zr+Zr) collisions, where the minimum bias trigger requires hits of both VPDs.
The collision centrality was determined from the measured multiplicity
of charged particles within $|\eta|<0.5$ compared to a Monte Carlo Glauber simulation~\cite{glauber,STAR:2021mii}.

The event plane angle $\Psi_n$ was determined by the tracks measured in the TPC, where $n$ denotes the harmonic order. 
The event plane resolution defined as $\langle\cos[n(\Psi_n^{\rm obs}-\Psi_n)]\rangle$~\cite{TwoSub} (``obs" indicates an observed angle) becomes largest around 10-30\% centrality ($\sim$0.62) for the second-order and 
at 0-5\% centrality ($\sim$0.38) for the third-order. 
Note that the perfect resolution corresponds to 1.0. 
The resolutions are very similar for the two isobar systems.
The event plane detector (EPD) located at forward and backward pseudorapidities (2.1$<|\eta|<$5.1) was also used for a cross check of the measurements, which provided consistent results with the TPC event plane measurements.
The results presented here utilize the TPC event plane measurements because of its superior resolution compared to the EPD ($\sim$0.38 (0.13) for the second-order (third-order) at the corresponding centralities).

To reconstruct \lam (\alam) hyperons, the decay channel of $\Lambda\rightarrow p\pi^-$ ($\bar{\Lambda}\rightarrow \bar{p}\pi^+$)
was utilized. The daughter charged tracks measured by the TPC were identified using the ionization energy loss in the TPC gas and
flight timing information from the TOF detector, and then \lam (\alam) hyperons were reconstructed
based on the invariant mass of the two daughters after applying cuts on decay topology to reduce combinatorial background.

Hyperon polarization is studied by utilizing parity-violating weak decays where the daughter baryon emission angle is correlated with the direction of the hyperon spin. The daughter baryon distribution in the hyperon rest frame can be written as:
\begin{eqnarray}
\frac{dN}{d\Omega^\ast} = \frac{1}{4\pi} (1+\alpha_H \textbf{\textit{P}}_H^\ast \cdot \hat{p}_B^\ast),
\end{eqnarray}
where $d\Omega^\ast $ is the solid angle element, and $\textbf{\textit{P}}_H^\ast$ and $\hat{p}_B^\ast$ denote hyperon polarization and the unit vector of daughter baryon momentum in the hyperon rest frame (as denoted by an asterisk); $\alpha_H$ is the hyperon decay parameter.
The decay parameter $\alpha_{\Lambda}$ for the decay $\Lambda\rightarrow p+\pi^-$ is set to $\alpha_\Lambda=0.732\pm0.014$~\cite{Zyla:2020zbs} assuming $\alpha_\Lambda=-\alpha_{\bar{\Lambda}}$. 
Polarization along the beam direction $P_z$~\cite{Adam:2019srw} is determined as
\begin{eqnarray}
P_z = \frac{\langle\cos\theta_p^\ast\rangle}{\alpha_H\langle\cos^2\theta_p^\ast\rangle},
\end{eqnarray}
where $\theta_p^\ast$ is the polar angle of the daughter proton in the \lam rest frame relative to the beam direction.  
The denominator $\langle\cos^2\theta_p^\ast\rangle$ accounts for the detector acceptance effect and is found to be close to 1/3, slightly depending on the hyperon's transverse momentum and centrality.

The systematic uncertainties were evaluated by variation of the topological cuts in the \lam reconstruction $\sim$3\% (10\%), using different methods of the signal extraction as explained  below $\sim$5\% (8\%), estimating possible background contribution to the signal $\sim$3\% (6\%), and uncertainty on the decay parameter $\sim$2\% (2\%).
The quoted numbers are examples of relative uncertainties for the second-order (third-order) results in 10-30\% (0-20\%) central collisions. All these contributions were added in quadrature, the value of which was quoted as the final systematic uncertainty.
The sine modulation of $P_z$ was extracted by measuring directly $\langle\cos\theta^\ast_p\sin[n(\phi-\Psi_n)]\rangle$ as a function of the invariant mass. 
The results were checked by measuring $\langle\cos\theta^\ast_p\rangle$, corrected for the acceptance effects, as a function of azimuthal angle relative to the event plane, fitting it with the sine Fourier function as presented below in Fig.~\ref{fig:costhe}, and followed by  correction for the event plane resolution (see Ref.~\cite{Adam:2019srw} for more details). 
It should be noted that $\langle\cos\theta^\ast_p\sin[n(\phi-\Psi_n)]\rangle$ can be directly calculated for a selected mass window if the purity of the \lam samples is high (the background contribution, if any, is negligible). The two approaches provide consistent results.
The EPD event plane and different sizes of TPC subevents (see Ref.~\cite{Adam:2019srw}) were also used for cross checks yielding consistent results as well. 
Self-correlation effects due to inclusion of the hyperon decay daughters in the TPC event plane determination were studied by excluding the daughters from the event plane calculation and ultimately found to be negligible.
The feed-down effect may dilute the $P_z$ sine modulation of primary $\Lambda$ by 10--15\%~\cite{Xia:2019fjf,Becattini:2019ntv} but since a correction for this effect is model-dependent, only results for inclusive $\Lambda$ are presented in this paper.

Figure~\ref{fig:costhe} shows $\langle\cos\theta_p^\ast\rangle^{\rm sub}$ as a function of \lam(\alam) azimuthal angle relative to the second- and third-order event planes, where the superscript ``sub" represents subtractions of the detector acceptance and inefficiency effects as described in Ref.~\cite{Adam:2019srw}. Furthermore, the results are multiplied by the sign of $\alpha_H$ for a clearer comparison between \lam and \alam. 
The right panel presents the measurement of the longitudinal component of polarization relative to the third-order event plane where sine patterns similar to those in the left panel are clearly seen, indicating the presence of triangular-flow-driven vorticity. It is noteworthy that while the origin of triangular flow is completely different than that of elliptic flow, a similar development of vorticity pattern is observed.
Since the results for \lam and \alam are consistent with each other, as expected in the vorticity-driven polarization picture (note that the difference observed in the third-order results is $\sim$1.4$\sigma$), both results are combined to enhance the statistical significance.
\begin{figure}[t]
\begin{center}\hspace{-0.5cm}
\includegraphics[width=0.51\linewidth]{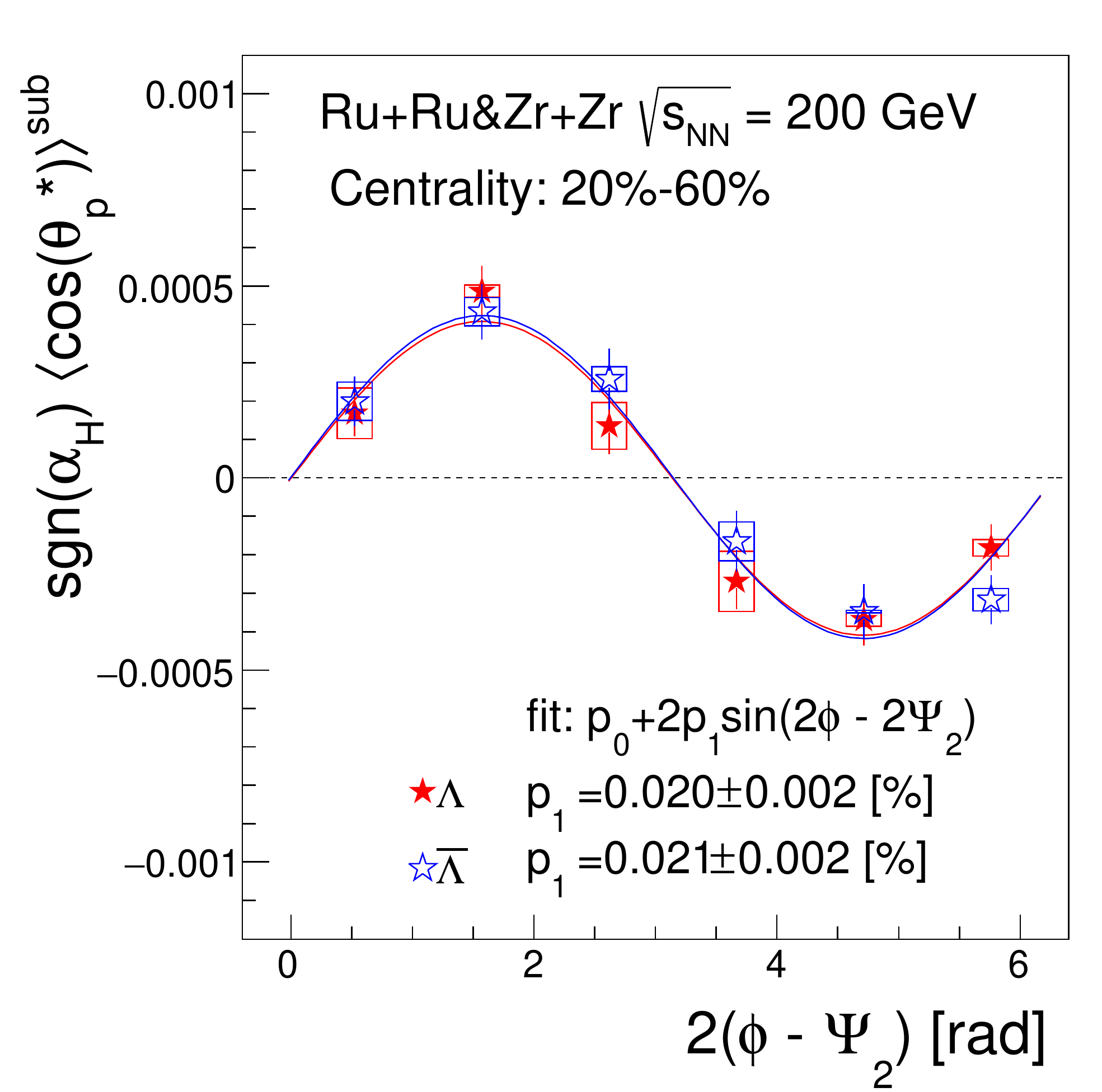}
\hspace{-0.2cm}
\includegraphics[width=0.51\linewidth]{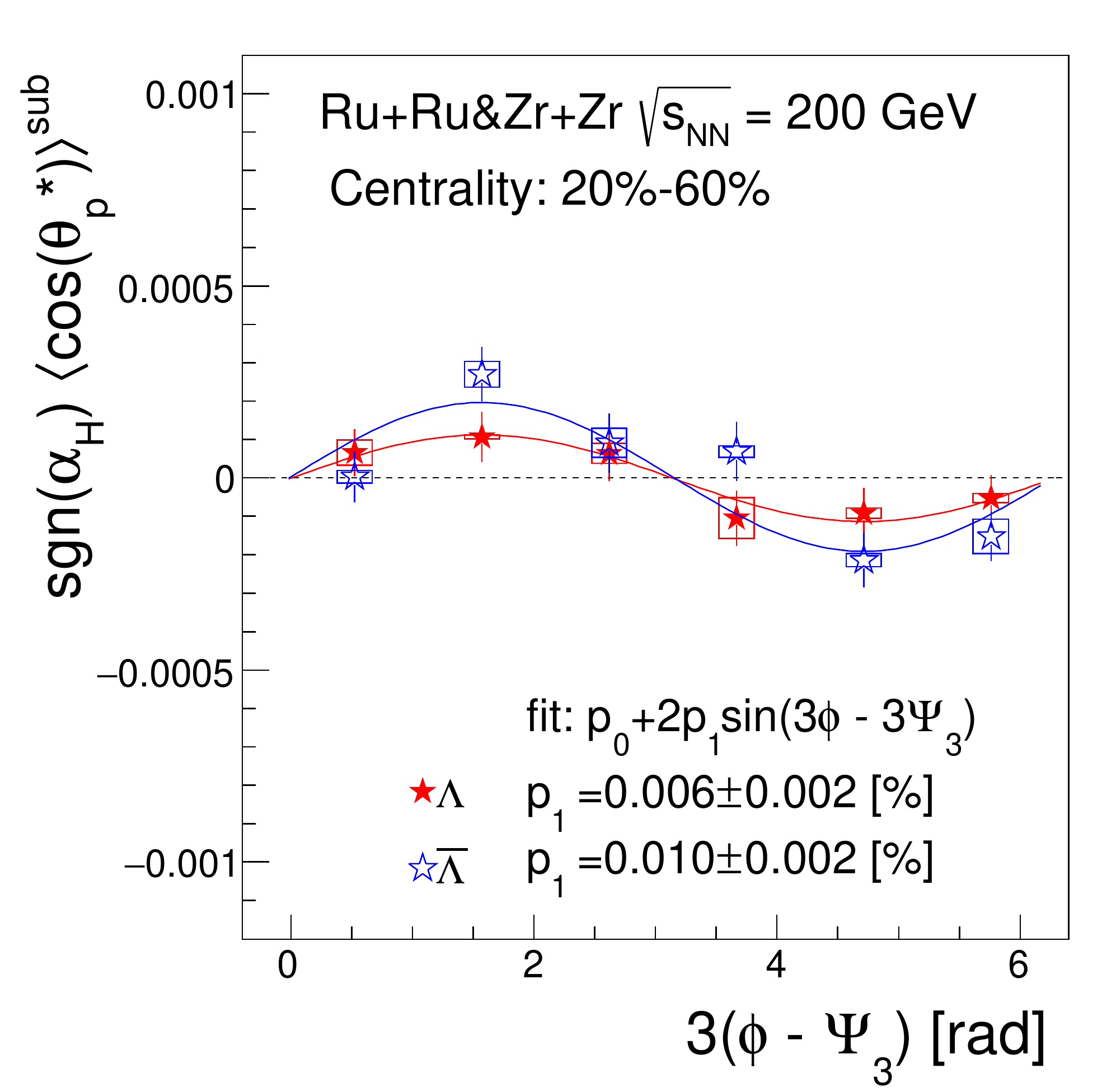}
\hspace{-0.5cm}
\caption{$\langle\cos\theta_p^\ast\rangle^{\rm sub}$ of \lam and \alam as a function of hyperon azimuthal angle relative to the second- (left panel) and the third-order (right panel) event planes, $n(\phi-\Psi_n)$, in 20-60\% central isobar collisions at \snn = 200 GeV. 
The sign of the data for \alam is flipped as indicated by ${\rm sgn}(\alpha_H)$. 
The solid lines are fit functions used to extract the parameters indicated in the label where $p_1$ corresponds to the $n^{\rm th}$-order Fourier sine coefficient. 
Note that the results presented in these figures are not corrected for the event plane resolution.}
\label{fig:costhe}
\end{center}
\end{figure}

\begin{figure}[thb]
\begin{center}
\includegraphics[width=\linewidth]{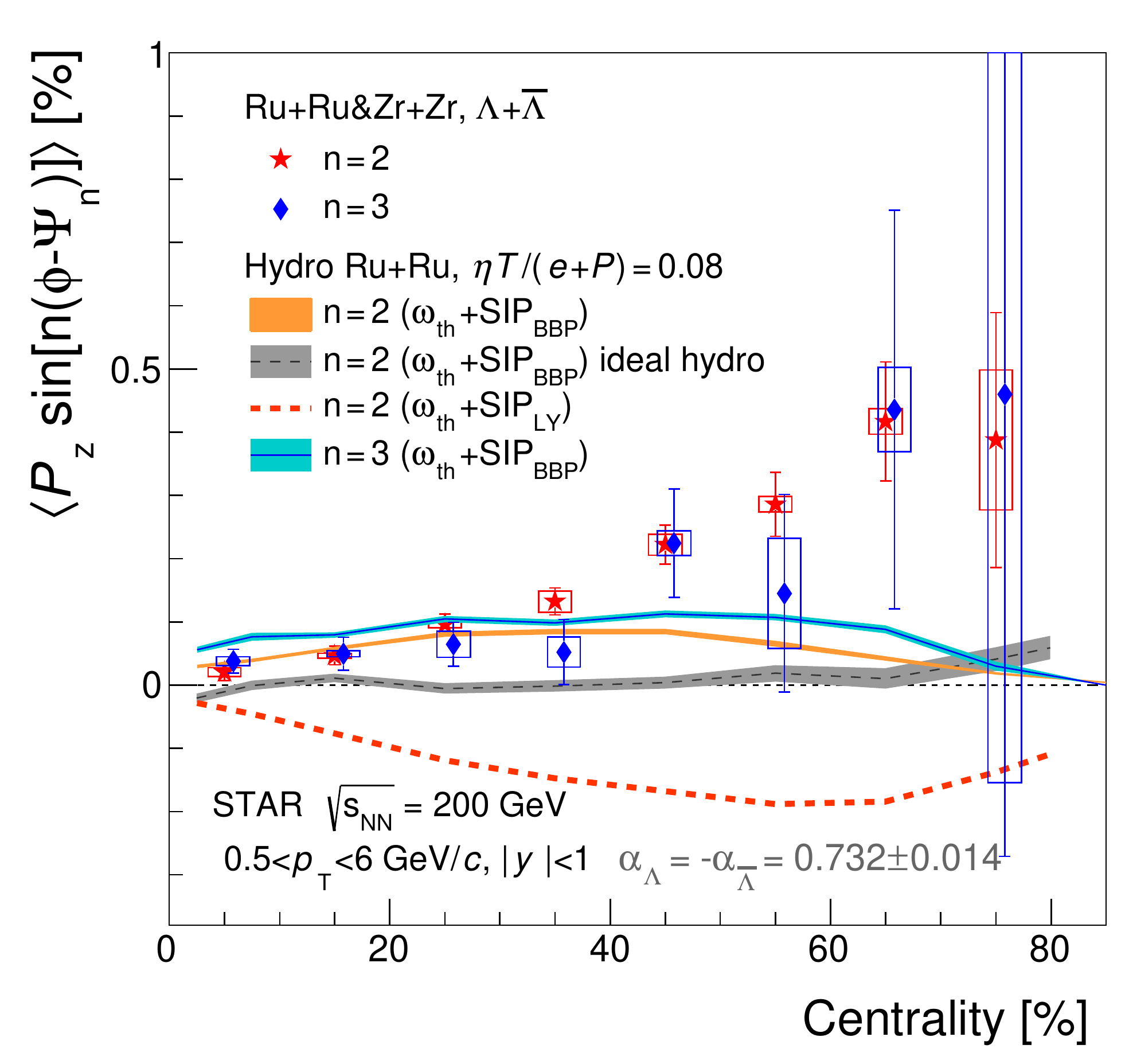}
\caption{Centrality dependence of the second- and the third-order Fourier sine coefficients of \lam+\alam polarization along the beam direction in isobar Ru+Ru and Zr+Zr collisions at \snn = 200 GeV. 
Open boxes show systematic uncertainties.
Solid bands show calculations from hydrodynamic model~\cite{Alzhrani:2022dpi} including contribution from the shear-induced polarization (SIP) based on Ref.~\cite{Becattini:2021suc} by Becattini-Buzzegoli-Palermo (BBP) or Ref.~\cite{Liu:2021uhn} by Liu-Yin (LY) in addition to that due to thermal vorticity $\omega_{\rm th}$. The model calculations with a nearly zero shear viscosity (``ideal hydro") are also shown.
}
\label{fig:PzvsCent}
\end{center}
\end{figure}

The sine modulations of $P_z$ are studied as a function of collision centrality and are presented in Fig.~\ref{fig:PzvsCent}.
Results of the measurements relative to both event planes are comparable in magnitude and exhibit similar centrality dependence, increasing in more peripheral collisions.
Calculations from a hydrodynamic model~\cite{Alzhrani:2022dpi} with specific shear viscosity $\eta T/(e+P)=0.08$ and including both the thermal vorticity and shear-induced contributions to the polarization are shown. The model results strongly depend on particular implementations of the shear-induced contribution.
The calculations with the shear contribution based on Ref.~\cite{Becattini:2021suc}, are in a rough qualitative agreement with the polarization sign and magnitudes for both harmonics, but fail to describe the data quantitatively especially in  peripheral collisions.
On the other hand, the calculation for the second-order with the shear contribution based on Ref.~\cite{Liu:2021uhn} shows the opposite sign to the data. Note that the model with Ref.~\cite{Liu:2021uhn} can provide the correct sign only if the $\Lambda$ mass is replaced with the mass of the constituent strange quark.

The model calculations with very small value of the specific shear viscosity $\eta T/(e+P)=0.001$ leads to almost zero $P_z$ as shown in Fig.~\ref{fig:PzvsCent}, indicating that the polarization measurements put an additional constraint on the shear viscosity values of the medium~\cite{Alzhrani:2022dpi}. Note that the hydrodynamic model calculations without the shear-induced polarization contribution always predict polarization with the opposite sign to that observed in the data. 

\begin{figure}[t]
\begin{center}
\includegraphics[width=\linewidth]{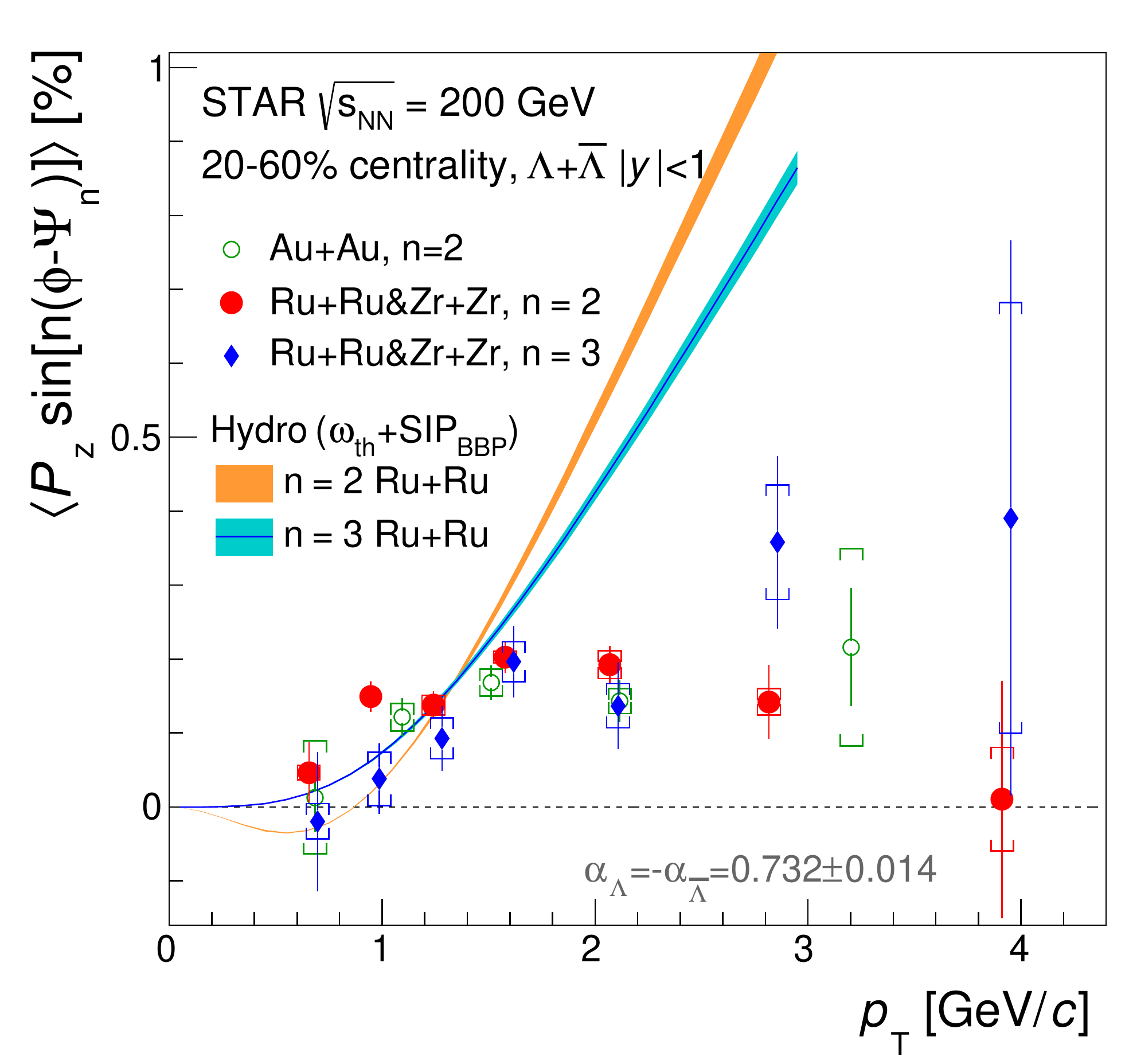}
\caption{Transverse momentum dependence of the second- and third-order Fourier sine coefficients of \lam+\alam polarization along the beam direction 
for 20-60\% central isobar Ru+Ru and Zr+Zr collisions at \snn = 200 GeV, compared to the second-order measurements in Au+Au collisions~\cite{Adam:2019srw}. 
Open boxes show systematic uncertainties. 
The results for the third-order event plane measurements in isobar collisions are slightly shifted for a better visibility. Solid bands present calculations from the hydrodynamic model~\cite{Alzhrani:2022dpi} (see Fig.~\ref{fig:PzvsCent} caption).
}
\label{fig:PzvsPT}
\end{center}
\end{figure}

If the observed polarization along the beam direction is induced by collective anisotropic flow, one might naively expect a transverse momentum dependence similar to that of the flow.
The $P_z$ sine modulations for measurements relative to both event planes are plotted as a function of hyperons' transverse momentum in Fig.~\ref{fig:PzvsPT}.
Results show that $p_T$ dependence of the polarization is indeed similar to that of elliptic ($v_2$) and triangular ($v_3$) flow. 
While the third-order $P_z$ modulation is smaller than the second-order for $p_T < 1.5$ GeV/$c$, the third-order results seem to increase faster, with a hint of out-pacing the second-order results at $p_T>2$ GeV/$c$. The significance of the third-order results away from zero is 4.8$\sigma$ for $1.1<\pt<6.0$ GeV/$c$ considering statistical and systematic uncertainties in quadrature.
A similar pattern is also observed in the flow measurements~\cite{ALICE:2016cti,PHENIX:2014uik} which further supports that the observed polarization is driven by collective flow. 
The hydrodynamic model calculations exhibit stronger $p_T$ dependence than that in the data and predict smaller values of the second-order polarization compared to the third-order at low $p_T$. In the model, such behavior is determined by two competing mechanisms, the thermal vorticity and the shear-induced polarization.
The second-order polarization results for isobar collisions are found to be comparable to or slightly higher than those for Au+Au collisions.

\begin{figure}[t]
\begin{center}
\includegraphics[width=\linewidth]{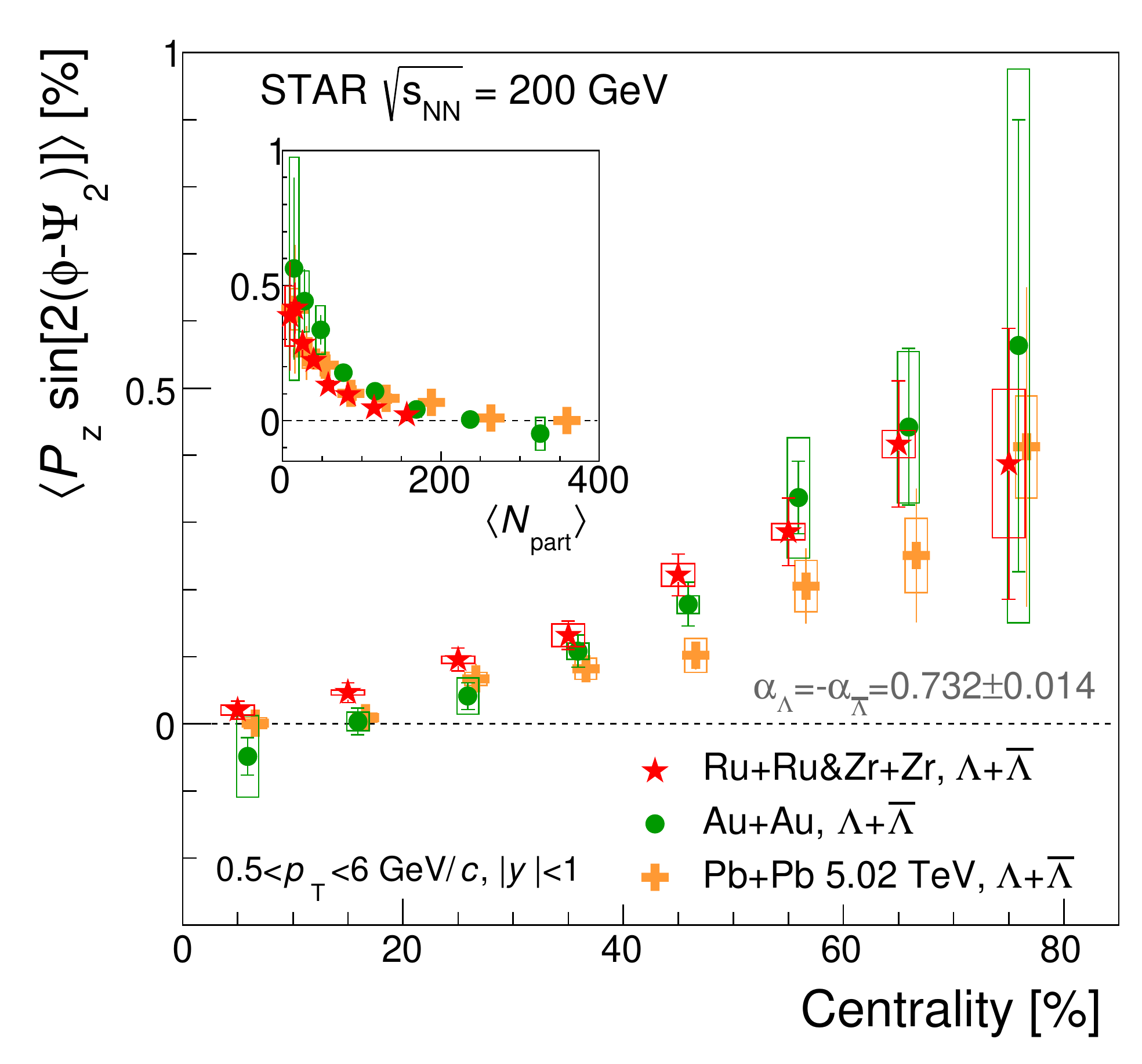}
\caption{Comparison of the second Fourier sine coefficients of \lam+\alam polarization component along the beam direction 
among isobar and Au+Au collisions at \snn = 200 GeV~\cite{Adam:2019srw} and Pb+Pb collisions at \snn = 5.02 TeV~\cite{ALICE:2021pzu} as a function of centrality. 
Open boxes show systematic uncertainties. 
The inset presents the same data plotted as a function of average number of participants $\langle N_{\rm part}\rangle$. Note that the data points for Pb+Pb collisions are rescaled to account for the difference in the decay parameter $\alpha_\Lambda$ used in Pb+Pb analysis.
}
\label{fig:Pz2vsPbPb}
\end{center}
\end{figure}

Figure~\ref{fig:Pz2vsPbPb} shows the centrality dependence of the second-order sine Fourier coefficients of $P_z$ in isobar collisions compared to results from Au+Au collisions at \snn = 200 GeV~\cite{Adam:2019srw} and Pb+Pb collisions at \snn = 5.02 TeV from the ALICE experiment~\cite{ALICE:2021pzu}.
The results do not show any strong energy dependence nor system size dependence for a given centrality.
The isobar collisions, a smaller system compared to Au+Au, show slightly larger polarization values in midcentral collisions, but the difference is not significant.
Note that the elliptic flow $v_2$ in 5.02 TeV Pb+Pb collisions~\cite{ALICE:2016ccg} is $\sim$60\% larger than that in 200 GeV isobar collisions~\cite{STAR:2021mii}. 
The data do not follow a naive expectation from the $v_2$ magnitude, i.e., larger local polarization in Pb+Pb for a given centrality.
The data are also plotted as a function of an average number of nucleon participants $N_{\rm part}$ estimated from the Glauber model in the inset of Fig.~\ref{fig:Pz2vsPbPb}, showing that the data scales better with $N_{\rm part}$, indicating a possible importance of the system size in vorticity formation.

In conclusion, \lam and \alam hyperon polarization along the beam direction has been measured in isobar Ru+Ru and Zr+Zr collisions at \snn = 200 GeV, with respect to the second-order event plane and, for the first time, to the third-order event plane.
The polarization is found to have a sinusoidal azimuthal dependence relative to both the event planes, indicating the creation of complex vorticity pattern induced by the elliptic and triangular flow in heavy-ion collisions.
The second- and third-order sine Fourier coefficients of the polarization exhibit increasing trends toward peripheral collisions and a mild \pt dependence similar to those of elliptic and triangular flow coefficients. 
Hydrodynamic model calculations including both thermal vorticity and thermal shear contributions based on ``BBP" implementation, qualitatively agree with the data predicting the correct sign for both harmonics, but underestimate the data in peripheral collisions and predict different shape of the \pt dependence. 
All other model calculations are in qualitative disagreement with our measurement. 
Comparison of the second-harmonic sine coefficient to those measured in 200 GeV Au+Au and 5.02 TeV Pb+Pb collisions, shows little system size and collision energy dependence of the polarization.
These results provide new insights into polarization mechanism and vorticity fields in heavy-ion collisions as well as additional constraints on properties and dynamics of the matter created in the collisions.

%
%
%
%
%
\begin{acknowledgments}
We thank the RHIC Operations Group and RCF at BNL, the NERSC Center at LBNL, and the Open Science Grid consortium for providing resources and support.  This work was supported in part by the Office of Nuclear Physics within the U.S. DOE Office of Science, the U.S. National Science Foundation, National Natural Science Foundation of China, Chinese Academy of Science, the Ministry of Science and Technology of China and the Chinese Ministry of Education, the Higher Education Sprout Project by Ministry of Education at NCKU, the National Research Foundation of Korea, Czech Science Foundation and Ministry of Education, Youth and Sports of the Czech Republic, Hungarian National Research, Development and Innovation Office, New National Excellency Programme of the Hungarian Ministry of Human Capacities, Department of Atomic Energy and Department of Science and Technology of the Government of India, the National Science Centre and WUT ID-UB of Poland, the Ministry of Science, Education and Sports of the Republic of Croatia, German Bundesministerium f\"ur Bildung, Wissenschaft, Forschung and Technologie (BMBF), Helmholtz Association, Ministry of Education, Culture, Sports, Science, and Technology (MEXT) and Japan Society for the Promotion of Science (JSPS).

\end{acknowledgments}
%
\bibliography{ref_isoPz}   
\end{document}